\documentstyle[12pt]{article}
\date{}
\begin{document}

\title{ Baryon Magnetic Moments and Axial Coupling Constants
with Relativistic and Exchange Current Effects}

\author{K.Dannbom$^1$, L. Ya. Glozman $^2$, C. Helminen $^1$
and D. O. Riska$^1$}
\maketitle

\centerline{\it $^1$Department of Physics, University of Helsinki,
00014 Finland}

\centerline{\it $^2$Institute for 
Theoretical Physics, University of Graz,
8010 Austria}

\vspace{1cm}

\centerline{\bf Abstract}
\vspace{0.5cm}

The large relativistic corrections to the constituent quark current
operators improve the predictions for the axial couplings of the
baryons, but worsen those for their magnetic moments. The 
exchange current corrections that are associated with 
flavor and spin dependent hyperfine interactions between the 
quarks with a form suggested by pseudoscalar meson exchange can 
compensate the
relativistic corrections to the baryon magnetic moments. This is
demonstrated by a calculation of the magnetic moments of the
non-strange and strange baryons using wave functions and exchange
current operators, which correspond to a recent phenomenological 
spin- and flavor
dependent 
interquark interaction model with a linear confining interaction, 
which yields a spectrum close to the
empirical one. The possibility that part of the flavor and spin
dependent interaction could be due to vector and axial-vector exchange
is explored.\\

\newpage

{\bf 1. Introduction}
\vspace{0.5cm}

As the constituent quarks are much lighter than the nucleon, both their
electromagnetic and the
axial current operators have significant
relativistic corrections, if the constituent quarks are treated as Dirac
particles \cite{Cl}. While the magnitude of these relativistic
corrections depends on the average velocity of the confined quarks and
thus on the model for the hyperfine interaction
between the quarks \cite{HaIs}, their effect is to reduce the
magnitude of the predicted values of both the axial coupling constants
and the magnetic moments of the baryons that are given by the static quark
model. While this correction reduces the standard overprediction of the
axial current coupling constants of the nucleons ($5/3$ vs $1.24$) 
and the strange
hyperons it worsens the mostly satisfactory predictions
for the magnetic moments of the baryons that are obtained with the
static quark model in the impulse approximation. 
We shall show here that the exchange current
correction that is associated with spin- and flavor dependent
interactions between the quarks, with the same operator structure as
that of the pseudoscalar meson octet exchange
interaction between the quarks can compensate for the relativistic
correction in the latter case, while leaving it operative in the case
of the axial coupling constants. This then suggests a way to obtain
an at least qualitatively satisfactory simultaneous description of both
the magnetic moments and the axial coupling constants although a
quantitative description would require a completely relativistic
quantum mechanical approach.\\

There are several 
different dynamical mechanisms that can give rise to flavor
dependent spin-spin interactions between constituent quarks
with the form \cite{GlRi}

$$-V(r_{ij})\vec{\lambda_i^F} \cdot \vec{\lambda_j^F} 
\vec{\sigma_i}\cdot\vec{\sigma_j}, \eqno(1.1)$$

\noindent
where the potential function $V(r_{ij})$ is positive at short range.
To these belong the interaction mediated by 
pseudoscalar and vector flavor-octet exchanges between the quarks 
\cite{GlRi}. In both these cases the correct sign of the interaction
(1.1) stems from the contact part of these interactions. This contact
interaction is opposite in sign to the long-range Yukawa tail.
In addition to these mechanisms there could be contributions 
of axial-vector
flavor-octet exchange interactions. In the latter case, in contrast,
the required negative sign in (1.1) is present in
the Yukawa tail and the contact
term is absent. The potential function $V(r_{ij})$ and its strength
can be determined phenomenologically by fitting the baryon spectrum.
However there exist a number of alternative and quite different 
parametrizations of $V(r_{ij})$ \cite{GPP, GPPVW} which 
yield predictions of approximately
the same quality for the baryon spectra. 
Hence the spectrum by itself cannot determine
the relative importance of the possible mechanisms above.
The coupling of the light pseudoscalar meson
octet, which forms the octet of Goldstone bosons of the spontaneously
broken approximate chiral symmetry of QCD, to the constituent quarks 
is probably a significant factor in the explanation of the spectra of
baryons and it also resolves
several of the problems with the naive quark model and the measured
spin- and strangeness content of the proton \cite{ChLi}. For the
present issue the salient point is that the flavor dependent octet
boson exchange interaction, by the requirement of current conservation,
implies the presence of octet vector two-body exchange magnetic moment
operators. As these increase the magnitudes of the magnetic moments of
the baryons that are predicted in the static quark model
\cite{BuHeYa}, they in principle should counteract the reduction
caused by the relativistic correction. There are no corresponding pure
pseudoscalar octet exchange contributions to the axial current of the
baryons.\\

We shall here construct the pseudoscalar boson exchange current so
as to be consistent with and to satisfy the continuity equation with
the pseudoscalar boson exchange interaction even when this is modified
phenomenologically at short
range. In the calculations we employ the explicit phenomenological
model for the hyperfine interaction between the quarks given in
ref. \cite{GPP}, which 
yields a satisfactory description of the nucleon and
$\Delta$-spectra in combination with a static linear confining
interaction. As 
the volume integral of that model for the hyperfine interaction does not
vanish, only part of it can be interpreted as being due to
pseudoscalar exchange mechanisms. The remaining part is treated 
phenomenologically,
using an appropriate exchange current operator for an interaction
with nonvanishing volume integral. \\

The construction of the flavor octet
exchange current operator is non-relativistic. That relativistic 
corrections to the
exchange current operator also are important is however evident from
the fact that the octet vector exchange currents, which are the quark
level analogues of the two-nucleon isovector exchange currents, are
inoperative in the case of the baryon decuplet, the flavor states of
which by definition are symmetric. In order to compensate
for the large relativistic corrections to the magnetic moment of the
decuplet baryons, as e.g. the $\Omega^-$, we therefore also consider
the lowest order relativistic corrections to the exchange
magnetic moment operators, 
which have flavor symmetric terms. These are the
quark level analogues of the (small) isoscalar two-nucleon 
exchange current operators, and arise from excitation of virtual
$q\bar q$-pairs. \\

In the 
present calculation we employ for the proton and the neutron the 3
quark wave function obtained in ref. \cite{GPP} by solving the Faddeev
equations with a spin- and flavor dependent model for the hyperfine
interaction between the quarks and a linear confining potential, which
yields a baryon spectrum that is close to the empirical one. For
the strange hyperons we use a phenomenological 
wave function model \cite{GlRi},
which is fitted to the numerically exact one for the nucleons,
 but which contains the masses of the constituent quarks as
explicit parameters and therefore permits extrapolation to the case of
baryons with strange quarks. As the quark Hamiltonian is purely
nonrelativistic we have to treat the required relativistic corrections
to the exchange current operators in an ad hoc way, which is suggested
by the relativistic corrections
that appear in the single quark current operators. The
present results therefore remain qualitative, but even so we
find that the exchange current contributions that are implied by the
quark-quark interaction model in ref. \cite{GPP} are more than
sufficient to compensate the reduction of the impulse approximation
values for the magnetic moments, which are caused by the relativistic
corrections. This is the main result of the present study. It
suggests that a unified description of the baryon axial and magnetic
constants can be achieved.
\\

The exchange current contributions that are associated with the flavor
and spin dependent hyperfine interaction model lead to overpredictions
of most of the baryon magnetic moments. These overpredictions can be
substantially reduced by the exchange current contribution that may be
associated with the confining interaction, under the assumption that
this can be effectively viewed as a (relativistic) scalar exchange interaction
\cite{BuHeYa}. In the cases where the former exchange current
contributions are too small for compensating the relativistic
corrections, as e.g. the $\Sigma^-$ and $\Omega^-$ hyperons, the
exchange current associated with the confining interaction does,
however, increase the
underprediction. Even so most
of the calculated magnetic moment values are very close to the
corresponding empirical values.\\

This paper is divided into 7 sections. In section 2 we discuss the
relativistic corrections to the single baryon current operators. The
static pseudoscalar exchange current magnetic moment operator
is derived in section 3. In
section 4 we derive the lowest order relativistic correction
to the pseudoscalar exchange current operators, which includes
flavor symmetric components. In section 5 we consider the exchange
current corrections that are associated with the part of the
phenomenological interaction, which cannot be interpreted as being due
to pseudoscalar exchange mechanisms. The magnetic moment
operator that may be
associated with the confining interaction is described in section 6.
A summarizing discussion is given in section 7. \\

\newpage

{\bf 2. Relativistic constituent quark currents}
\vspace{0.5cm}

Under the assumption that the constituent quarks can be treated as
point Dirac particles without anomalous terms, their electromagnetic
and axial current operators are

$$<p'|J_\mu(0)|p>=ie\bar u(p')\gamma_\mu[{1\over 2}\lambda_3+{1\over
2\sqrt{3}}\lambda_8]u(p),\eqno(2.1a)$$
$$<p'|A_{\mu a}(0)|p>=ig_A^q\bar u(p')\gamma_\mu\gamma_5{\lambda_a\over
2}u(p).\eqno(2.1b)$$
Here $g_A^q$ is the axial current coupling of the constituent quarks,
which in the large color limit is 1 \cite{We}, and with inclusion of
the lowest $1/N_c$ 
correction is $g_A^q\simeq 0.87$ \cite{Wei}. An argument
for the absence of anomalous terms in the electromagnetic
current (2.1a) has been given in ref. \cite{Di}.\\

In terms of Pauli spinors these operators reduce to

$$\vec J=e[{1\over 2}\lambda_3+{1\over 2\sqrt{3}}\lambda_8]
\frac{1}{\sqrt{1+\vec v^2}}\{[\vec
v+i{\vec \sigma \times \vec q\over 2m_q}]$$
$$-
\frac{\sqrt{m_q^2+(\vec P+\vec q/2)^2}
-\sqrt{m_q^2+(\vec P-\vec q/2)^2}}
{2m_q[1+\sqrt
{1+\vec v^2}]}[{\vec q\over
2m_q}+i\vec \sigma
\times \vec v]\} ,\eqno(2.2a)$$
$$\vec A_a=-g_A^q\frac{\lambda_a}{2}\vec\sigma
\{1-{2\over 3}(1-{1\over \sqrt{1+\vec v^2}})\}.\eqno(2.2b) $$
Here the velocity 
operator is defined as $\vec v=\vec P/m_q$, 
with $\vec P=(\vec p'+\vec p)/2$ and $m_q$ is the mass 
of the constituent quark. The momentum transfer is denoted
$\vec q=\vec p'-\vec p$. Note that in the expression for the
electromagnetic current (2.2a) the limit $\vec p=\vec p'$ has been
taken in all terms, where it does not affect the derivation of the
magnetic moment operator $\vec \mu\equiv-{i\over 2}(\nabla \times
\vec q)_{\vec q=0}$. The spin term in the
magnetic moment operator that corresponds to the electromagnetic
current operator (2.2a) is then
$$\vec \mu_{spin}={e\over 2m_q}[{1\over 2}\lambda_3+{1\over
2\sqrt{3}}\lambda_8]\frac{\vec
\sigma}{\sqrt{1+\vec v^2}}\{1-{1\over 3}(1-\frac{1}{\sqrt{
1+\vec v^2}})\}.\eqno(2.3)$$

To lowest order in $\vec v^2$ the expressions for the spin
magnetic moment and axial current operators are then

$$\vec \mu_{spin}={e\over 2m_q}[{1\over 2}\lambda_3+{1\over
2\sqrt{3}}\lambda_8]\vec \sigma(1-{2\vec v^2\over 3}), 
\eqno(2.4a) $$
$$\vec A_a = -g_A^q \frac{\lambda_a}{2}\vec\sigma
(1-\frac{\vec v^2}{3}).\eqno(2.4b)$$
The relativistic $\vec v^2$ correction is large in the case of the
constituent quarks, the masses of which are of the order 300-400 MeV.
Its order of
magnitude is easiest to estimate with the harmonic oscillator model
for the quark confinement, in
which case 

$$<\vec v^2>={\omega\over m_q}={1\over m_q^2r^2},\eqno(2.5)$$
where $\omega$ is the oscillator parameter and $r$ the mean radius ($r
= 1/\sqrt{m_q\omega}$). As $r\sim 0.86$ fm for the baryons, the latter
expression (2.5) gives $<\vec v^2>=0.45$ if $m_q=$ 340 MeV, which is an
appropriate value for the nonstrange constituent quarks. This shows that
the relativistic correction to the magnetic moments amounts to
a large reduction of the values predicted by the
static quark model. The large value of $<\vec v^2>$ also shows that
the $v/c$ expansion is unreliable, and that the unapproximated
expressions (2.2b) and (2.3) have to be employed.
With $m_u=340 $ MeV the empirical radius $0.86 $ fm corresponds to
$\omega=154$ MeV in the harmonic oscillator model. This is an 
appropriate value for the unperturbed zero-order wave function
under the assumption that the hyperfine
interaction (1.1)
can be treated in first order perturbation theory.
When the 3-quark system is solved with a linear confining potential
and with full account of the pseudoscalar exchange interaction to all
orders the wave function is much more compact, with a matter $rms$
radius of $0.47$ fm \cite{GPP}. In the oscillator model this
corresponds to an oscillator frequency of $\sim 540$ MeV (2.5) (
here this oscillator frequency effectively includes effects of both
confining and hyperfine interactions). We shall here employ the known
exact wave function for this model for the non-strange baryons
\cite{GPP} and an oscillator model wave function with this
oscillator frequency for the strange hyperons, as this should provide
the most appropriate extrapolation to the states with the heavier
strange constituent quarks for which we use the mass 460 MeV
\cite{GlRi}. \\

Consider first the magnetic moments of the ground state baryons. In
the impulse approximation the static quark model expressions for the
magnetic moments are linear combinations of the ratios of the nucleon
to the up, down and strange constituent quark masses (Table 1). With
the usual approximation that the constituent masses of the up and down
quarks are equal, the two ratios to be considered are $m_N/m_u$ and
$m_N/m_s$. In Table 1 we have taken the relativistic correction
in the expression (2.3) for the magnetic moment operator
into account by the replacements \cite{Br}

$${1\over m}\rightarrow {1\over m^*}=
{1\over m}<{1\over \sqrt{1+\vec
v^2}}(1-{1\over 3}(1-{1\over \sqrt{1+\vec v^2}}))>,\eqno(2.6)$$
in the quark masses in the impulse approximation expressions.\\

The results in column I in Table I show that all the measured baryon
magnetic moments, with exception of the $\Delta^{++}$ are
satisfactorily ($\pm 15\%$) predicted in the static impulse
approximation in the absence of the relativistic correction. 
The calculation of the neutron
and proton magnetic moments was done with
the exact 3 quark wavefunction of ref. \cite{GPP}. The same
relativistic reduction would in the oscillator model be obtained with
a wave function with 
an "effective" oscillator frequency of 540 MeV. In the absence of a
similar exact wave function for the hyperons  we have therefore
used the oscillator model with $\omega_{eff}=$ 540 MeV to estimate the
relativistic corrections to the magnetic moments of the strange
hyperons and 
the $\Delta$:s. Including the relativistic correction reduces the
predicted magnetic moment values by 20--30\%, and notably worsens the
agreement with the experimental values. We shall show
in the following section that the exchange current corrections can
compensate these reductions in most cases.\\

In Table 2 the axial vector couplings of the octet baryons are listed.
With the exception of $g_A(\Xi^-\rightarrow \Sigma^0$) the empirical
values are 
considerably smaller in magnitude than the values predicted by the
static quark model. The static quark model predictions are most
conveniently given as the $D,F$ coefficient values
$$D=1,\quad F={2\over 3}.\eqno(2.7)$$
The corresponding empirical values are 0.77 and 0.45 \cite{GaSa},
which are 23\% and 32\% smaller than the static quark model values
respectively. The static quark model values as obtained with $g_A^q=1$
are given in column II of Table 2. The corresponding values that are
obtained with the relativistic correction in eq. (2.2b) taken into
account are listed in column III. The exact wave function of ref.
\cite{GPP} was used for the calculation of the relativistic
correction to $g_A(n\rightarrow p)$. In the case of the strange
hyperons we again employed the oscillator wave function model with
$\omega=$ 540 MeV, which leads to the same result as the exact wave
function in the case of $g_A(n\rightarrow p)$.
Here we have used the average of the
constituent masses of the light and strange quarks in the evaluation
of the relativistic correction for strangeness changing decays $(\bar
m_q=\sqrt{340\cdot460}$ MeV $=395$ MeV). 
In this case the relativistic correction, which amounts
to a reduction of $\sim$ 19\%, leads to
notably better overall agreement with the empirical values, with
exception of the case of $g_A(\Xi^-\rightarrow \Sigma^0$). In column
IV we show the results obtained with $g_A^q=0.87$ \cite{Di} and
inclusion of the relativistic correction. These results are in very
good agreement with the empirical values.\\

\vspace{1cm}

{\bf 3. The pseudoscalar exchange magnetic moment operator}
\vspace{0.5cm}

The nonrelativistic pseudoscalar octet exchange interaction 
between two constituent
quarks has the following general expression in momentum space:
$$V(\vec k)=-v(\vec k)\vec \sigma^1\cdot \vec k \vec \sigma^2\cdot
\vec k\vec \lambda^1\cdot \vec \lambda^2,\eqno(3.1)$$
in the $SU(3)_F$ symmetric limit. Here $\vec k$ is the momentum
transfer and $v(\vec k)$ a scalar potential function
and the superscripts refer to the quark number. The exchange
current operator, which satisfies the continuity equation with this
interaction, yields - to lowest order in $v/c$ - the magnetic moment
operator \cite{GlRi,TsRiBl}:
$$\vec M_{ex}={1\over 2}\{(\vec \tau^1\times
\vec \tau^2)_3+\lambda^1_4\lambda_5^2-\lambda^1_5\lambda_4^2\}$$
$$\{v(k)\vec \sigma^1\times \vec \sigma^2-{1\over k}{\partial
v(k)\over \partial k}\vec k\times (\vec k\times (\vec \sigma^1\times
\vec \sigma^2))\},\eqno(3.2)$$
to which also should be added a term proportional to the
center-of-mass coordinate, but which does not contribute to $S$-state
quarks.\\

For the ground state baryons, we shall only need to consider the
spatial scalar components of the interaction (3.1) and
exchange magnetic moment operator (3.2), which are
$$V_S(\vec k)=-{1\over 3}v(k)k^2\vec \sigma^1\cdot \vec \sigma^2\vec
\lambda^1\cdot \vec \lambda^2,\eqno(3.3a)$$
$$\vec M_S={1\over 2}\{(\vec \tau^1\times
\vec\tau^2)_3+\lambda_4^1\lambda_5^2-\lambda_5^1\lambda_4^2\}$$
$$\{v(k)+{2\over 3}k{\partial v(k)\over \partial k}\}\vec
\sigma^1\times \vec \sigma^2.\eqno(3.3b)$$
Fourier transformation to configuration space yields
$$V_S(\vec  r)={1\over 3}f(r)\vec \sigma^1\cdot \vec \sigma^2\vec
\lambda^1\cdot \vec \lambda^2,\eqno(3.4a)$$
$$\vec M_S=-{1\over 2}g(r)\{(\vec \tau^1\times
\vec\tau^2)_3+\lambda_4^1\lambda_5^2-\lambda^1_5\lambda_4^2\}
\vec \sigma^1\times \vec \sigma^2,\eqno(3.4b)$$
where we have used the notation
$$f(r)=\nabla^2\tilde{v}(r),\eqno(3.5a)$$
$$g(r)=\tilde{v}(r)+{2\over 3}\vec r\cdot \vec
\nabla\tilde{v}(r),\eqno(3.5b)$$
and where $\tilde{v}(r)$ is 
the Fourier transform of the potential function
$v(k)$:
$$\tilde{v}(r)={1\over 2\pi ^2r}\int_{0}^{\infty}dkk
\sin(kr)v(k).\eqno(3.6)$$
By way of illustration we note that for a bare pseudoscalar meson
exchange interaction the potential function $v(k)$ is 
$$v(k)={g^2\over 4m_1m_2}{1\over \mu^2+k^2}\eqno(3.7)$$
where $m_1$ and $m_2$ are the constituent masses of the two
interacting quarks, $\mu$ is the meson mass and $g$ is the meson-quark
coupling constant. \\

Insertion of this expression in (3.5) and (3.6) yields the following
expressions for the functions $f(r)$ and $g(r)$: 
$$f(r)={g^2\over 4\pi}{1\over 4m_1m_2}\{\mu^2{e^{-\mu
r}\over r}-4\pi\delta(\vec r)\}\eqno(3.8a)$$
$$g(r)=-{g^2\over 4\pi}{\mu\over 12m_1m_2}(2\mu r-1){e^{-\mu r}\over
\mu r}.\eqno(3.8b)$$
These bare pseudoscalar exchange operators do not take into account 
the spatially
extended structure of the constituent quarks and the
pseudoscalar mesons 
and should be modified
accordingly at short distances prior to comparison with data. \\

The volume integral of the function $f(r)$ in (3.8a) vanishes. This is a
consequence of the fact that the coupling of pseudoscalar mesons to
quarks vanishes with momentum, as implied by the expression
(3.3a), by which $V_S(\vec k=0)=0$ for mesons with finite mass. 
As $V_S(\vec k=0)$ is proportional
to the volume integral of $f(r)$, the latter also vanishes.\\

In practice the potential function $f(r)$ will be determined by fits
to the baryon spectra. One then needs an algorithm for determining the
function $g(r)$, given $f(r)$. This is obtained by solving (3.5a) for
$\tilde{v}(r)$ and then inserting the result in (3.5b). The sought for
expression for $g(r)$ is then
$$g(r)=-{1\over 3}\{2\int_{r}^{\infty}dr'r'f(r')-{1\over
r}\int_{r}^{\infty}dr'\int_{r'}^{\infty}dr''r''f(r'')\}.\eqno(3.9)$$
It is readily seen that insertion of the example function (3.8a) in
this expression yields the correct answer (3.8b) for $g(r)$. If the
volume integral of the phenomenologically determined model function
$f(r)$ does not vanish an inconsistency in the expressions for the
interaction and exchange current operators above appears, however. 
This implies that part of the effective spin-flavor interaction (3.4a)
would not have its origin in pseudoscalar exchange
mechanisms, but in exchange mechanisms
of shorter range as e.g. axial vector exchange, which does not
contribute to the exchange magnetic moment in lowest order.
The corresponding exchange magnetic moment operator (3.4b) constructed
with $g(r)$ determined by (3.9) in this case has to be renormalized
down by the
fraction of the phenomenological interaction that could be due to 
axial vector exchange,
which does not give rise to any exchange moment to lowest order 
\cite{TsRiBl}.\\

In ref. \cite{GPP} it has been shown that a very satisfactory
prediction for the spectra of the nucleon and the $\Delta$-resonance is
obtained if the confining potential between the quarks is taken to be
linear $(0.474$ fm$^{-2} r)$ and the function $f(r)$ in the pseudoscalar
exchange interaction (3.4a) is taken to have the form
$$f(r)={g^2\over 4\pi}{1\over 4m_1m_2}\{\mu^2{e^{-\mu r}\over
r}H(r)-{4\over
\sqrt{\pi}}\alpha^3e^{-\alpha^2(r-r_0)^2}\}.\eqno(3.10)$$
Here $H(r)$ is a function of the form 
$$H(r)=\{1-{1\over 1+e^{\beta(r-r_0)}}\}^5,\eqno(3.11)$$
which cuts off the Yukawa function at $r_0=0.43$ fm. The value of the
parameter $\beta$ is 20 fm$^{-1}$.
In (3.10) the
$\delta$-function has been smeared over a range $\alpha^{-1}$, where
$\alpha =2.91$ fm$^{-1}$, which corresponds to the radius of the
constituent quarks (the interaction model of ref. [4]
also contains a smaller spin dependent but 
flavor independent term, which will not be considered here). 
The $\pi$-quark coupling constant $g$ is $g^2/4\pi
=0.67$, which corresponds to the value $g_{\pi NN}^{2}/4\pi=14.2$
\cite{GlRi}. The volume integral of the function $f(r)$ does not
vanish. The ratio of the volume integral of the negative inner and the
positive outer part of $f(r) (f(1.26$ fm)=0) is 10.6. This suggests
that only the fraction $X=1/10.6 \simeq 0.094$ of the short range part
of the interaction should be interpreted as arising from 
pseudoscalar meson exchange (the outer
part has the Yukawa tail, which corresponds to pion exchange).
We shall therefore multiply the calculated exchange magnetic moments
with this factor $X$. The remaining fraction $1-X=0.906$ of the
negative part of the phenomenological interaction (3.10) could be interpreted
as being due to axial vector exchange mechanisms for which there is 
no requirement
on the volume integral, and for which exchange current
corrections appear 
only as relativistic corrections. These will be treated
in the following subsection.
We shall show that when $g(r)$ is calculated from the
expression (3.9), with the model function (3.10) renormalized as
mentioned above (using for
simplicity  the unit step function $\theta(r-r_0)$ for $H(r)$), the
corresponding exchange current correction to the magnetic moments of
the baryons go in the right direction for compensating the
relativistic corrections, 
except in the case of the decuplet baryons that have
symmetric flavor 
wave functions and those octet baryons for which all the 3
quarks have equal charge as the $\Sigma^-$ and $\Xi^-$. \\

In Table 3 we list the matrix elements of the flavor-spin operators
$(F,S)_1$ $=(\vec\tau^1\times \vec
\tau^2)_3(\vec \sigma^1\times \vec \sigma^2)$ and
$(F,S)_2=(\lambda^1_4\lambda^2_5-\lambda^1_5\lambda_4^2)
(\vec \sigma^1\times
\vec \sigma^2)_3$ in the exchange magnetic moment operator (3.2) for
the octet baryons and the $\Delta\rightarrow N$ transition magnetic
moment. Because of the antisymmetry of the flavor parts of these
operators they have no matrix element for the baryon decuplet. \\

The contributions to the baryon magnetic moments of the exchange
current operator (3.4b) can then be expressed (in units of nuclear
magnetons) as
$$\mu^{ex}=-{1\over 2m_N}m_N<g(r)_{g.s.}(F,S)>=
-{1\over 2m_N}m_N<g(r)>_{g.s.}<(F,S)>,\eqno(3.12)$$
where $m_N$ is the nucleon mass, $<g(r)>_{g.s}$ is the spatial matrix
element of the function $g(r)$ for the ground state and the matrix
elements of the flavor-spin operator $(F,S)$ are those given in Table 3.
The first expression for $\mu^{ex}$ in eq. (3.12) should be used with
the 3 quark wave function of ref. \cite{GPP}, while the latter expression
in eq. (3.12) can be used with the oscillator model. Since $g(r)$
is dependent on the quark masses $m_1$ and $m_2$ (eqs. (3.9) and 
(3.10)), the appropriate 
masses should be taken into account when calculating
$<g(r)>_{g.s.}$. The masses $m_1=m_2=340$ MeV are used together with the 
flavor-spin matrix elements $<(F,S)_1>$, while the masses $m_1=340$ MeV 
and $m_2=460$ 
MeV should be used with $<(F,S)_2>$.\\

For the bare pion exchange potential (3.7) the matrix element
$<g(r)>_{g.s.}$ is very small and positive, which leads to a small
negative
value for this exchange current contribution to the proton magnetic
moment. The short range modification of the meson exchange potential
leads to a change of sign of this contribution \cite{BuHeYa}. 
The pseudoscalar exchange potential with the short range modification
(3.10) leads to 
values for the matrix element $m_N<g(r)>_{g.s.}$ in (3.12)
which are strongly dependent on the parameter $r_0$ in the short range
modification (3.10) as well as on the size of the baryon wave function.
With $\omega_{eff}=$ 540 MeV in the oscillator model, and using the
same parameter values in the pseudoscalar potential (3.10) as in ref.
\cite{GPP}, we obtain $m_N<g(r)>_{g.s.}=0.056$ (after multiplication
by the renormalization factor $X=0.094$). This implies an exchange
current correction to the proton
magnetic moment of 0.22 n.m. 
With the exact wave function of ref. \cite{GPP} the
value of the exchange current correction is slightly larger (0.23 n.m.).
As this exchange current contribution was
calculated without account of relativistic corrections, it is expected
to represent an overestimate.\\

The main relativistic correction to the pseudoscalar exchange
magnetic moment operator (3.4b) can be inferred from the
expansion to order $1/m_q^4$ of the general pseudoscalar
exchange magnetic moment operator in ref. \cite{TsRiBl}. This 
correction
can to a good approximation be taken into account as an
overall correction factor $(1-4 \vec v^2/3)$. As the origin of
this factor is the spinor normalization factor and the energy
denominator in the small components of the quark wave functions,
it is most properly taken into account in unexpanded form
as an overall factor $(1+4 \vec v^2/3)$ in the 
denominator of the expression
(3.4b) for the exchange current operator. This correction
factor reduces the estimate of the pseudoscalar exchange
contribution to the magnetic moment of the proton in the oscillator 
model by a factor 0.46 from 0.22 n.m. to the smaller value 0.10 n.m.\\

The exchange current corrections
to the magnetic moments of the octet baryons calculated and
renormalized in this way 
are listed in column "EXCI" in Table 4.\\

\vspace{1cm}

{\bf 4. The flavor symmetric pseudoscalar exchange current}
\vspace{0.5cm}

The flavor antisymmetric exchange magnetic moment operator (3.2)
considered above is associated with the static nonrelativistic
pseudoscalar exchange interaction (3.1). When the exchange current
operator that is associated with the relativistic form of the
pseudoscalar exchange interaction is expanded in powers of $v/c$,
flavor symmetric terms appear in the order $m_q^{-4}$. The
corresponding magnetic moment operator has been given for the case of
the two-nucleon system in ref. \cite{TsRiBl}. Applying the results of
ref. \cite{TsRiBl} to the case of two quarks that interact by exchange
of the $SU(3)_F$ octet of pseudoscalar mesons yields an exchange
magnetic moment operator that can be decomposed into flavor symmetric
and flavor antisymmetric parts as
$$\vec M_{ex}=\vec M_{ex}^{F,S}+\vec M_{ex}^{F,A}.\eqno(4.1)$$
The flavor symmetric exchange magnetic moment operator can be written
as
$$\vec M_{ex}^{F,S}={1\over 8m_1m_2}v(k)(\vec \sigma^1+\vec
\sigma^2)\cdot \vec k\vec k$$
$$[{2\over 3}\vec \tau^1\cdot \vec \tau^2+
(\vec\tau^1+\vec\tau^2)_3+{2\over
3}(\lambda^1_4\lambda_4^2+\lambda^1_5\lambda^2_5)$$
$$-{4\over 3}(\lambda^1_6\lambda^2_6+\lambda^1_7\lambda^2_7)-{2\over
3}\lambda^1_8\lambda^2_8$$
$$+{2\over 3\sqrt{3}}(\lambda^1_8+\lambda^2_8)+{1\over
\sqrt{3}}(\lambda^1_3\lambda^2_8+\lambda^1_8\lambda^2_3)],\eqno(4.2)$$
where $v(k)$ is the pseudoscalar interaction potential defined in
(3.1). The flavor antisymmetric term is correspondingly
$$\vec M_{ex}^{F,A}={1\over 8m_1m_2}v(k)(\vec \sigma^1-\vec
\sigma^2)\cdot \vec k\vec k$$
$$[(\vec\tau^1-\vec\tau^2)_3+{2\over 
3\sqrt{3}}(\lambda^1_8-\lambda^2_8)$$
$$+{1\over
\sqrt{3}}(\lambda^1_8\lambda^2_3-\lambda^1_3\lambda^2_8)],\eqno(4.3)$$
in addition to a term with the same flavor operator as the main flavor
antisymmetric exchange magnetic moment operator (3.2), and which may
be viewed as a relativistic correction to it. The importance of the
flavor symmetric exchange magnetic moment operator (4.2) derives from
the fact that it gives a nonvanishing contribution to the magnetic
moments of the decuplet baryons as well as the $\Sigma^-$ and $\Xi^-$
octet hyperons, to which the flavor antisymmetric terms cannot
contribute.\\

In the case of the ground state baryons, which have all constituent
quarks in the lowest $S$-state, only the spatial scalar components of
the spin operators in (4.2) and (4.3) contribute. By retaining only
that and taking the Fourier transform yields the exchange magnetic
moment operators
$$\vec M_{ex}^{F,S}=-{1\over 24m_1m_2}f(r)(\vec \sigma^1+\vec
\sigma^2)_3[{2\over 3}\vec 
\tau^1\cdot \vec \tau^2+(\vec\tau^1+\vec\tau^2)_3$$
$$+{2\over 3}(\lambda^1_4\lambda^2_4+\lambda^1_5\lambda^2_5)$$
$$-{4\over 3}(\lambda^1_6\lambda^2_6+\lambda^1_7\lambda^2_7)-{2\over
3}\lambda^1_8\lambda^2_8$$
$$+{2\over 3\sqrt{3}}(\lambda^1_8+\lambda^2_8)+{1\over
\sqrt{3}}(\lambda^1_3\lambda^2_8+\lambda^1_8\lambda^2_3)],\eqno(4.4a)$$

$$\vec M_{ex}^{F,A}=-{1\over 24m_1m_2}f(r)(\vec \sigma^1-\vec
\sigma^2)_3[(\vec \tau^1-\vec\tau^2)_3$$
$$+{2\over 3\sqrt{3}}(\lambda^1_8-\lambda^2_8)+{1\over
\sqrt{3}}(\lambda^1_8\lambda^2_3-\lambda^1_3\lambda^2_8)].\eqno(4.4b)$$
Here the function $f(r)$ is the Laplacian of the pseudoscalar exchange
interaction, as defined in eq. (3.5a). Note that the factor ${1\over
m_1m_2}$ in eqs. (4.4a) and (4.4b) is not the same in all parts of 
the expressions. In the calculations we will use the masses $m_1=m_2=340$
MeV for the non-strange baryons and $m_1=m_2=460$ MeV for the $\Omega^-$ 
hyperon.
In all other cases we make a decomposition of the expressions for 
$\vec M_{ex}^{F,S}$ and $\vec M_{ex}^{F,A}$ so that terms with flavor
operators containing the isospin operators $\vec\tau$ have $m_1=m_2=340$
MeV and other terms have
$m_1=340$ MeV, $m_2=460$ MeV. When using the oscillator model 
the calculation of 
$<\vec M_{ex}^{F,S}>$ and $<\vec M_{ex}^{F,A}>$ can be simplified as
described below.\\

The fact that the 
radial behaviour of the exchange magnetic moment operators
is directly proportional 
to the pseudoscalar exchange interaction implies
that the corresponding magnetic moments of the ground state baryons
may for the oscillator model be expressed in terms of the orbital 
matrix element of the
pseudoscalar exchange interaction for the ground state baryons. If the
whole effective interaction (3.4a) were due to pseudoscalar exchange,
this matrix element would, as defined in ref. \cite{GlRi}, be
$$P_{00}=-{1\over 3}<f(r)>_{g.s.}.\eqno(4.5)$$
In first order perturbation theory the matrix element $P_{00}$ may be
determined directly from the empirical $N-\Delta$ splitting, which in
that approximation is simply $10 P_{00}$, so that $P_{00}\simeq$ 29
MeV \cite{GlRi}. With the model (3.10) for the radial shape of the
fine structure interaction the matrix element 
calculated with the oscillator model ($\omega_{eff}=540$ MeV) 
is as large as 
101 MeV. Note that only the fraction $X=0.094$ of this matrix
element can - as explained above - be ascribed to pseudoscalar
exchange. The relativistic correction to the proton exchange
magnetic moment would 
then be 0.15 n.m. (0.17 n.m. with the exact wave function
of ref. \cite{GPP}). If as in section 3 above we include a factor
$(1+4\vec v^2 /3)$ in the denominators of the expressions
(4.2) and (4.3) (or alternatively in the expression (4.5))
to account for relativistic corrections from the Dirac
operators this estimate for
the matrix element $P_{00}$ in the oscillator model
drops to only 46 MeV. This value for $P_{00}$ has still to be
renormalized down, however, 
in view of the fact that the volume integral of
the phenomenological potential function (3.10) does not vanish, as
would be required by pseudoscalar exchange mechanisms. For
the $K$ meson exchange contribution the corresponding matrix
element is calculated as in eq. (4.5), but with the average mass 
$\bar m_q=395$ MeV. When including the factor ${1/(1+4\vec v^2 /3)}$,
also calculated with $\bar m_q$, one gets 41 MeV.
Denoting these two matrix elements $P_{00}^\pi$ and
$P_{00}^K$ and assuming that the $\eta$ meson exchange contribution is 
$P_{00}^\eta=P_{00}^K$ one may express the orbital matrix elements of
the exchange magnetic moments (4.4) as
$$<\vec M_{ex}^{F,S}>=\mu_NX({m_N\over 4m_1m_2})(\vec \sigma^1+\vec
\sigma^2)_3$$
$$\{[{2\over 3}\vec \tau^1\cdot \vec
\tau^2+(\vec \tau^1+\vec \tau^2)_3]P_{00}^\pi$$
$$+[{2\over
3}(\lambda^1_4\lambda^2_4+\lambda^1_5\lambda^2_5)-{4\over
3}(\lambda^1_6\lambda^2_6+\lambda^1_7\lambda^2_7)]P_{00}^K$$
$$-[{2\over 3}\lambda^1_8\lambda^2_8-{2\over
3\sqrt{3}}(\lambda^1_8+\lambda^2_8)-{1\over
\sqrt{3}}(\lambda^1_3\lambda^2_8+\lambda^1_8\lambda^2_3)]P_{00}^\eta\}
,\eqno(4.6a)$$

$$<\vec M_{ex}^{F,A}>=\mu_NX({m_N\over 4m_1m_2})(\vec \sigma^1-\vec
\sigma^2)_3.$$
$$\{(\vec \tau^1-\vec \tau^2)_3P_{00}^\pi$$
$$+{1\over \sqrt{3}}[{2\over
3}(\lambda^1_8-\lambda^2_8)+(\lambda^1_8\lambda^2_3-\lambda^1_3
\lambda^2_8)]P_{00}^\eta\}.\eqno(4.6b)$$
The matrix elements of the flavor-spin operators in these two
expressions 
are listed in Table 5. The factor $X$ has the value 0.094. \\

The contributions to the baryon magnetic moments of the exchange
current operators (4.6) can be expressed as
$\mu_{rel}^{ex}=\mu_N(m_N/4m_1m_2)$ times the sum of the flavor-spin
matrix elements in Table 5. As in eq (4.4) the masses $m_1$ and
$m_2$ should be taken as $m_1=m_2=340$ MeV for non-strange baryons and
for strange baryons when associated with the matrix element $P_{00}^\pi$,
as $m_1=m_2=460$ MeV for the $\Omega^-$ hyperon and otherwise as
$m_1=340$ MeV and $m_2=460$ MeV. 
The numerical values obtained in this way for the pseudoscalar
exchange magnetic moments of order $m^{-4}$  
are given in column EXCII in Table 4. \\

The results in Table 4 show that the typical magnitude of the
contributions to the baryon magnetic moments of the relativistic
corrections to the exchange current operators are somewhat smaller than
that of the static exchange current operator (3.12) considered in the
previous section. They also in
every instance go in the right direction to improve the agreement
between the quark model predictions and the empirical values for the
magnetic moment operators. In the case of the decuplet baryons 
it is only
the relativistic exchange current that 
gives a contribution and thus it is important for the compensation 
of the relativistic corrections.\\

\vspace{1cm}

{\bf 5. Exchange magnetic moment for phenomenological short range interactions}
\vspace{0.5cm}

As the volume integral of the phenomenological interaction model
(3.10) \cite{GPP} does not vanish, a fraction $(1-X)$ of the short
range part of it has to be interpreted as arising from other than
pseudoscalar exchange mechanisms.\\

If the phenomenological quark-quark interaction is expressed in terms
of the usual 5 Fermi invariants, only the tensor $(T)$ and axial
vector $(A)$ invariants give rise to spin-spin interactions of the
form (3.4a), that have non-vanishing volume integrals in lowest order
in $1/m$ \cite{BR}. Since the form of these interactions and the
corresponding exchange current operators are very similar, we restrict
the consideration here to the axial vector invariant.\\

While the spin-spin interactions, which are associated with the $T$
and $A$ invariants, are of order $m^0$, the corresponding exchange
magnetic moment operators are of order $m^{-2}$ - i.e. they have the
form of relativistic corrections. If the spin-spin interaction has the
form (3.4a), and is assumed to arise from an interaction with the
operator structure of the $A$ invariant, the contribution
to lowest 
order in $1/m$ to the corresponding exchange magnetic moment is
$$\vec M_{ex}(A)=-{(1-X)\over 24m_1m_2}f(r)\{(\vec \sigma^1+\vec
\sigma^2)_3[{2\over 3}\vec \tau^1\cdot \vec \tau^2+(\vec \tau^1+\vec
\tau^2)_3$$
$$+{2\over 3}(\lambda^1_4\lambda^2_4+\lambda^1_5\lambda^2_5)$$
$$-{4\over 3}(\lambda^1_6\lambda^2_6+\lambda^1_7\lambda^2_7)-{2\over
3}\lambda^1_8\lambda^2_8$$
$$+{2\over
3\sqrt{3}}(\lambda^1_8+\lambda^2_8)+{1\over
\sqrt{3}}(\lambda^1_3\lambda^2_8+\lambda^1_8\lambda^2_3)]$$
$$+(\vec \sigma^1-\vec \sigma^2)_3[(\vec \tau^1-\vec \tau^2)_3+{2\over
3\sqrt{3}}(\lambda^1_8-\lambda^2_8)$$
$$+{1\over \sqrt{3}}(\lambda^1_8\lambda^2_3-\lambda^1_3\lambda^2_8)]$$
$$-2(\vec \sigma^1\times \vec \sigma^2)[(\vec \tau^1\times
\vec \tau^2)_3+\lambda^1_4\lambda^2_5-\lambda^1_5\lambda^2_4]\}.
\eqno(5.1)$$
From this expression a small term involving the derivative of $f(r)$
has been left out. With exception for the last term in the bracket, 
this operator is
identical in form to the relativistic pseudoscalar exchange operators
(4.4a) and (4.4b). The overall factor $1-X=0.906$ gives the fraction
of the short range part of the phenomenological interaction (3.10),
which could be interpreted as arising from axial vector exchange
mechanisms. The expression (5.1) for the magnetic moment operator
associated with the axial vector invariant was obtained by
generalizing the corresponding $SU(2)$ expressions given in ref.
\cite{TsRiBl} to $SU(3)$ and Fourier transforming the resulting
expression.\\

In view of the similarity in form between this expression and the
expressions (4.4), its matrix elements for the ground state can be
obtained by the same method as used in going from the expressions
(4.4) to the corresponding matrix element expressions (4.6). In this
case the indices $\pi,K$ and $\eta$ on the radial integrals $P_{00}$
(4.6) only indicate that the integrals are assumed to depend on the
quark masses in the same way as the matrix elements of $\pi, K$ and
$\eta$ exchange interactions. The contributions to the baryon magnetic
moments given by the magnetic moment expressions (5.1) calculated in
this way and using the same numerical values as in the expressions
(4.4) are listed in the column EXCIII in Table 4. The required matrix
elements of the spin-flavor operators are given in Tables 3 and 5.\\

It is worth noting, that while the pseudoscalar exchange mechanisms
give rise to an axial exchange current operator, which is only of order
$m^{-5}$, axial exchange mechanisms give rise to exchange currents of
order $m^{-3}$ \cite{TsRi}. While the contributions of the former
represent 4th order relativistic corrections, and thus should be
insignificant, the latter may give rise to non-negligible contributions
to the axial coupling constants of the baryons. A quantitative
calculation of these would require a completely covariant framework,
and is therefore not attempted here.\\

\vspace{1cm}

{\bf 6. The Confinement Current}
\vspace{0.5cm}

If a spin- and flavor independent confining interaction is formally
viewed as the static approximation to a relativistic scalar exchange
interaction, with positive instead of the conventional negative sign,
an exchange current of order $m^{-2}$ may be associated with it
\cite{BuHeYa}. The form of the associated magnetic moment operator
will then be
$$\vec M_{ex}(C)=-{e\over 2m_1m_2}\{[{1\over 2}\lambda_3^1+{1\over
2\sqrt{3}}\lambda^1_8]\vec \sigma^1$$
$$+[{1\over 2}\lambda^2_3+{1\over 2\sqrt{3}}\lambda^2_8]\vec
\sigma^2\}v_C(r),\eqno(6.1)$$
where $v_C(r)$ is the static confining interaction. With a wave
function model, in which the spatial and the flavor-spin component
factorize -- as e.g. the harmonic oscillator model -- in the
case of quarks with equal mass this operator may
be rewritten in the simpler form
$$\vec M_{ex}(C)=-{e\over m^2}v_C(r_{12})[{1\over
2}\lambda^1_3+{1\over 2\sqrt{3}}\lambda^1_8]\vec \sigma^1,
\eqno(6.2)$$
where $r_{12}=|\vec r_1-\vec r_2|$. This implies that it can be viewed
as a correction to be included along with the single nucleon magnetic
moment operators. For the ground state wave function with 3 $s$-state
quarks it can then be included by multiplication of the magnetic
moment expressions by a factor $F$ defined as 
$$F\equiv 1-2{<v_C(r_{12})>\over m}\eqno(6.3)$$

In the interaction model of ref. \cite{GPP} the confining interaction
is taken to be the linear potential $v_C(r_{12})=c r_{12}$, with
$c=0.474$ fm$^{-2}$. With oscillator model wave functions this leads
to the following explicit expressions for the factor $F$:
$$F=1-{2c\over m}\sqrt{{8\over \pi m\omega}}.\eqno(6.4)$$
The correction term in this expression gives a contribution $-1.12$
n.m. to the predicted magnetic moment of the proton. Introduction of
the same relativistic correction factor that is assumed to arise from
the Dirac spinors in a nonrelativistic reduction of the relativistic
scalar ("S") invariant
as was used in the exchange current expressions above, 
reduces this correction by a factor 0.46, and the net
correction to the magnetic moment of the proton due to this 
model for the exchange magnetic
moment associated with the confinement current is then --0.51 n.m. 
If this
correction is subtracted from the sum of the other exchange
current contributions in Table 4 one obtains the prediction
$\mu_p=2.81$ n.m., which is close to the empirical value 2.79 n.m.\\

In Table 4 the contributions from the exchange magnetic moment
operator (6.1) are listed in the column
CONF. In the numerical values the proper masses of the quark pair in
the two-body matrix elements involved have been taken into account.
These corrections bring the predicted magnetic moment values
closer to the empirical values in most cases, the only exceptions
being the $\Sigma^-$, $\Xi^-$ and $\Omega^-$ hyperons.\\

\vspace{1cm}

{\bf 7. Discussion}
\vspace{0.5cm}

The results in Tables 1 and 4 show that the corrections to the baryon
magnetic moments that arise from exchange
current operators are large, when using the phenomenological
interaction model of ref. \cite{GPP}, which leads to a very
satisfactory baryon spectrum, and that they tend to 
compensate the large relativistic
corrections to the static quark model predictions for the magnetic
moments of the baryons. The absence of pure pseudoscalar exchange
currents of low order to the axial current of the baryons on the other
hand implies that the corresponding if somewhat smaller relativistic
corrections to the axial coupling constants of the baryons remain
uncompensated (Table 2).\\
 
The very large exchange current contributions that were found to be
associated with the phenomenological flavor-spin interaction (3.10)
were found to lead to large overpredictions of the baryon magnetic
moments. The relativistic corrections that arise from the Dirac spinors
in the nonrelativistic reduction of the exchange current operators
reduce their matrix elements by about one half. To reduce the
remaining - and still considerable - overprediction of the magnetic
moments the exchange current contribution that would be associated with
the confining interaction, under the assumption that this can be formally
viewed as a scalar exchange interaction, was invoked. 
This exchange current -
first derived in ref. \cite{BuHeYa} - is also large, and brings about
the desired cancellation of the overpredictions,
and to what in the end are mostly satisfactory
magnetic moment predictions. If the effective confining 
interaction is interpreted as
a vector exchange interaction the numerical values
would be very similar. 
In the case of the $\Sigma^-$, $\Xi^-$
and $\Omega^-$ hyperons the magnetic moments are however
underpredicted. \\

This last problem indicates that the flavor and spin
structure of the exchange magnetic moment operator is not completely
adequate.  There is obviously a need
for a consistent relativistic treatment as well as more information on
the shorter range vector and axial exchange contributions to the
phenomenological interaction. 
It should be stressed that the meson 
exchange type description of the hyperfine interaction between the
constituent quarks, which leads to very satisfactory predictions for
all measured baryon spectra \cite{GlRi,GPP,GR} requires the
presence of these magnetic moment operators through the continuity
equation because of the flavor dependence of the interaction.
On the other hand the fact that the relativistic corrections
to the exchange current operator considered in section 4 are
only slightly smaller than the non-relativistic static
exchange current contributions considered in section 3
emphasizes the need for a fully relativistic treatment.\\

The treatment of the pseudoscalar octet exchange current contributions
to the magnetic moments of the baryons differs from previous work
\cite{BuHeYa,Ro} in that we consider the full octet exchange current
operator, and in the treatment of the short range part of the
pseudoscalar exchange interaction, and finally in that we also
consider the lowest order relativistic corrections to the exchange
current operator. In ref. \cite{BuHeYa} the exchange currents
associated with the one gluon exchange interaction between the
constituent quarks was also considered and played a very important role
there, but this contribution is likely to be much less
significant than what was suggested in ref. \cite{BuHeYa} in view of
the insignificance of the one gluon exchange interaction that is
indicated by the baryon spectrum \cite{GlRi}.\\

Finally it should be noted that a fraction of the phenomenological
interaction (3.4a) could also due to vector instead of pseudoscalar
and axial exchange mechanisms. The spin-spin component of the
vector exchange interaction has to have zero volume integral as does
the corresponding pseudoscalar model (3.10). In the case of
vector exchange mechanisms, the relation (3.5b) between the orbital
part of the magnetization density should be replaced by 
$$g(r)=2\tilde{v}(r)+\vec r\cdot \vec \nabla \tilde{v}(r).\eqno(7.1)$$
If this relation were used in place of the relation (3.5b) the
exchange current contributions in the column EXCI of Table 4 would be
somewhat reduced.\\

The separation of the exchange magnetic moment operator here into a
term that was associated with the pseudoscalar octet exchange
mechanism, and another purely phenomenological term was motivated by
the fact that the phenomenological interaction model developed in ref.
\cite{GPP} has a large volume integral, whereas it should vanish for a
pseudoscalar exchange interaction. Given only a
spin-flavor interaction of the form (1.1) does not, however,
permit a unique determination of the fraction of it that may be
ascribed to pseudoscalar meson exchange, nor does it indicate what
the dynamical origin of the short range terms that lead to the
nonvanishing volume integral is. Hence there remains a substantial
uncertainty as to the proper form of the exchange current operator, as
the continuity equation determines only its longitudinal part. In view of
this uncertainty, we view the predicted exchange current contributions
in Table 4 as no more than suggestive, and expect them to be very
strongly dependent on the particular form of the phenomenological
interaction employed here. As a concrete example of this, the volume
integral of the interaction (3.10) is very sensitive to the parameter
$r_0$ in (3.10) and can be reduced by 50\% by a small reduction of 
its value. The fact that the fraction of the phenomenological
interaction used here, which can be ascribed to pseudoscalar
exchange mechanisms, is very small is due to the fact that 
the interaction changes sign at the  
large radius 1.26 fm. With a smaller value of the radius where the
interaction changes sign, the volume integral decreases, and hence a
correspondingly larger fraction of the interaction can be 
ascribed to
pseudoscalar exchange mechanisms. For comparison
the corresponding isospin dependent spin-spin interaction 
component in a recently
developed realistic phenomenological nucleon-nucleon 
interaction model [21] changes sign already at 0.6 fm.\\

In view of this the numerical predictions above should
be viewed as suggestive rather than as definitive.
The numerical values depend
strongly on the particular parametrization of the potential
function in (1.1) and its interpretation as well as
on the very specific mechanisms
for the exchange currents considered here. Equally important is
that the nonrelativistic scheme considered here 
and use of the $v/c$ expansions for the exchange current
operators and the employment of nonrelativistic wave functions cannot 
be fully adequate as
$v/c \simeq 1$. The main conclusion of the present
work is therefore that a unified description of the axial constants 
and magnetic moments of the baryons appears to be possible.

\

\vspace{1cm}

{\bf Acknowledgement}
\vspace{0.5cm}

We thank Dr. Varga for supplying us with the convenient
parametrization of the nucleon wave function for the model
of ref \cite{GPP} and
Professor M. Rosina for an instructive discussion. 
DOR thanks Professor V. Vento for valuable correspondence.

\newpage

\centerline{\bf Table 1}
\vspace{0.5cm}

Magnetic moments of the ground state baryon octet and the
$\Delta^{++}$ and $\Omega^-$ 
(in nuclear magnetons). Column IA contains the quark model impulse 
approximation 
expressions with the relativistic corrections. Columns I and II
contain the impulse approximation values without  and with  the
relativistic correction. The empirical values are from ref. [11-13].\\

\begin{center}
\begin{tabular}{|l|l|r|r|r|} \hline
 & IA & exp & I & II \\ \hline
 &&&& \\
$p$ & ${m_N\over m_u^*}$ & +2.79 & +2.76 & 1.80 \\
 &&&& \\
$n$ & $-{2\over 3}{m_N\over m_u^*}$ & --1.91 & --1.84 & --1.20\\
 &&&& \\
$\Lambda$ & $-{1\over 3}{m_N\over m_s^*}$ & --0.61 & --0.67 & --0.48 \\
 &&&& \\
$\Sigma^+$ & ${8\over 9}{m_N\over m_u^*}+{1\over 9}{m_N\over m_s^*}$ &
+2.46 & +2.68 & 1.76 \\
 &&&& \\
$\Sigma^0$ & ${2\over 9}{m_N\over m_u^*}+{1\over 9}{m_N\over m_s^*}$ &
? & +0.84 & 0.56 \\
 &&&& \\
$\Sigma^0 \rightarrow \Lambda$ & $-{1\over \sqrt{3}}{m_N\over m_u^*}$
& $|1.61|$ & --1.59 & --1.04  \\
 &&&& \\
$\Sigma^-$ & $-{4\over 9}{m_N\over m_u^*}+{1\over 9}{m_N\over m_s^*}$
& --1.16 & --1.00 & --0.64  \\
 &&&& \\
$\Xi^0$ & $-{2\over 9}{m_N\over m_u^*}-{4\over 9}{m_N\over m_s^*}$ &
--1.25 & --1.51 & --1.04 \\
 &&&& \\
$\Xi^-$ & ${1\over 9}{m_N\over m_u^*}-{4\over 9}{m_N\over m_s^*}$ &
--0.65 & --0.59 & --0.44  \\ 
 &&&& \\ \hline
&&&& \\
$\Delta^{++}$ & $2{m_N\over m_u^*}$ & 4.52 & 5.52 & 3.60 \\
 &&&& \\
$\Delta^+\rightarrow p$ & ${2\sqrt{2}\over 3}{m_N\over m_u^*}$ & 3.1 &
2.6 & 1.70 \\
 &&&& \\
$\Omega^-$ & $-{m_N\over m_s^*}$ & --2.019 & --2.01 & --1.44  \\
 &&&& \\
\hline
\end{tabular}
\end{center}
\newpage
\centerline{\bf Table 2}
\vspace{0.5cm}

The axial coupling constants of the baryon octet. Column I gives the
expressions in terms of the F and D coefficients and column II the
static quark model prediction (with $g_A^q=1$). Column III gives the
predicted values with inclusion of the relativistic correction with
$g_A^q=1$ and column IV the predictions with $g_A^q=0.87$. The
empirical values are taken from refs. \cite{PaDaGr,GaSa}.\\

\begin{center}
\begin{tabular}{|l|l|r|r|r|r|} \hline
 & I & exp & II & III & IV \\ \hline
 &&&&& \\
$n\rightarrow p$ & $F+D$ & 1.26 & 1.67 & 1.35 & 1.17 \\
 &&&&& \\
$\Sigma^{\pm}\rightarrow \Lambda$ & $\sqrt{{2\over 3}}D$ & 0.62 & 0.81
& 0.66 & 0.57\\
 &&&&& \\
$\Sigma^- \rightarrow \Sigma^0$ & $\sqrt{2}F$ & 0.67 & 0.94 & 0.76 &
0.66 \\
 &&&&& \\
$\Lambda \rightarrow p$ & $-\sqrt{3\over 2}(F+{D\over 3})$ & 0.88 &
1.22 & 1.01 & 0.88 \\
 &&&&& \\
$\Sigma^-\rightarrow n$ & $-(F-D)$ & 0.34 & 0.33 & 0.28 & 0.24 \\
 &&&&& \\
$\Xi^-\rightarrow \Lambda$ & $\sqrt{{3\over 2}}(F-{D\over 3})$ & 0.31
& 0.40 & 0.34 & 0.30\\
 &&&&& \\
$\Xi^-\rightarrow \Sigma^0$ & ${1\over \sqrt{2}} (F+D)$ & 1.36 & 1.18 &
0.97 & 0.84 \\
 &&&&& \\
$\Xi^0\rightarrow \Sigma^+$ & $F+D$ & ? & 1.67 & 1.38 & 1.20\\
 &&&&& \\
$\Xi^- \rightarrow \Xi^0$ & $F-D$ & --0.28 & --0.33 & --0.27 & --0.23 \\
 &&&&& \\ \hline
\end{tabular}
\end{center}

\newpage

\centerline{\bf Table 3}
\vspace{0.5cm}

The matrix elements of the flavor-spin operators
$(F,S)_1=
(\vec\tau^1\times\vec\tau^2)_3(\vec 
\sigma^1\times \vec \sigma^2)_3$ and
$(F,S)_2=(\lambda^1_4\lambda^2_5-\lambda^1_5\lambda^2_4)
(\vec \sigma^1\times
\vec \sigma^2)_3$ for the ground state baryons.\\

\begin{center}
\begin{tabular}{|l|c|c|} \hline
 & $<(F,S)_1>$ &
$<(F,S)_2>$\\\hline
 && \\
$p$ & --4 & 0\\
 && \\
$n$ & 4 & 0\\
 && \\
$\Lambda$ & 0 & 2\\
 && \\
$\Sigma^+$ & 0 & --4\\
 && \\
$\Sigma^0$ & 0 & --2\\
 && \\
$\Sigma^0\rightarrow \Lambda$ & ${4\over \sqrt{3}}$ & ${2\over
\sqrt{3}}$\\
 && \\
$\Sigma^-$ & 0 & 0\\
 && \\
$\Xi^0$ & 0 & 4\\
 && \\
$\Xi^-$ & 0 & 0\\
 && \\\hline
 && \\
$\Delta^+\rightarrow p$ & $-4\sqrt{2}$ & 0\\
 && \\
 \hline
\end{tabular}
\end{center}
\newpage

\centerline{\bf Table 4}
\vspace{0.5cm}

The magnetic moments of the ground state baryon octet and the
$\Delta^{++}$ and the $\Omega^-$ (in nuclear magnetons). In column
I+R the impulse approximation values that include the relativistic
corrections are given. In column EXCI the contributions from the
nonrelativistic pseudoscalar exchange current operator (3.3b) are
given. Column EXCII contains the contribution from the relativistic
correction to the pseudoscalar exchange current operator (4.6). Column
EXCIII contains the contribution from the short range exchange current
operator (5.1). The column CONF contains the correction from the
exchange current associated with the confining interaction. 
The
sums of the exchange current and impulse approximation results are
listed in column Total. \\

\begin{center}
\begin{tabular}{|l|r|r|r|r|r|r|r|} \hline
 & I+R & EXCI & EXCII & EXCIII & CONF & Total & exp\\ \hline
 &&&&&&& \\
$p$ & 1.80 & 0.10 & 0.07 & 1.35 & --0.51 & 2.81 & 2.79\\
 &&&&&&& \\
$n$ & --1.20 & --0.10 & --0.04 & --1.06 & 0.34 &--2.06 & --1.91\\
 &&&&&&& \\
$\Lambda$ & --0.48 & --0.05 & --0.01 & --0.35 & 0.12 & --0.77 & --0.61\\
 &&&&&&& \\
$\Sigma^+$ & 1.76 & 0.09 & 0.06 & 0.99 & --0.44 & 2.46 & 2.46\\
 &&&&&&& \\
$\Sigma^0$ & 0.56 & 0.05 & 0.02 & 0.37 & --0.14 & 0.86 & ?\\
 &&&&&&& \\
$\Sigma^0\rightarrow \Lambda$ 
& --1.04 & --0.08 & --0.03 & --0.80 & 0.26 & --1.69 &
$ \vert 1.61 \vert$ \\
 &&&&&&& \\
$\Sigma^-$ & --0.64 & 0 & --0.03 & --0.25 & 0.16 & --0.76 & --1.16\\
 &&&&&&& \\
$\Xi^0$ & --1.04 & --0.09 & --0.03 & --0.69 & 0.23 & --1.62 & --1.25\\
 &&&&&&& \\
$\Xi^-$ & --0.44 & 0 & --0.02 & --0.20 & 0.10 & --0.56 & --0.65\\ \hline
 &&&&&&& \\
$\Delta^{++}$ & 3.60 & 0 & 0.18 & 1.76 & --1.02 & 4.52 & 4.52\\
 &&&&&&& \\
$\Delta^+\rightarrow p$ & 1.70 & 0.14 & 0.06 & 1.50 & --0.48 
& 2.92 & 3.1\\
 &&&&&&& \\
$\Omega^-$ & --1.44 & 0 & --0.05 & --0.44 & 0.28 & --1.65 & --2.019\\
 &&&&&&& \\ \hline
\end{tabular}
\end{center}

\newpage
\centerline{\bf Table 5}
\vspace{0.5cm}

The matrix elements of the combinations of symmetric and antisymmetric
flavor and spin operators defined in eqs. (4.6) for the ground state
baryons. The spin and flavor operators are defined as follows: $S_1=
(\vec \sigma^1+\vec\sigma^2)_3$,
$S_2=(\vec \sigma^1-\vec\sigma^2)_3$, $F_1^S={2\over 3}
\vec \tau^1\cdot \vec
\tau^2+(\vec \tau^1+\vec \tau^2)_3$, $F_2^S={2\over
3}(\lambda^1_4\lambda^2_4+\lambda^1_5\lambda^2_5)-{4\over
3}(\lambda^1_6\lambda^2_6+\lambda^1_7\lambda^2_7)$, $F_3^S=
-[{2\over 3}\lambda^1_8\lambda^2_8-{2\over3\sqrt{3}}
(\lambda^1_8+\lambda^2_8)-
{1\over\sqrt{3}}(\lambda^1_3\lambda^2_8+\lambda^1_8\lambda^2_3)]$, 
$F_1^A=
(\vec \tau^1-\vec \tau^2)_3$ and $F_2^A={1\over \sqrt{3}}[{2\over
3}(\lambda^1_8-\lambda^2_8)+(\lambda^1_8\lambda^2_3-\lambda^1_3
\lambda^2_8)]$.\\

\begin{center}
\begin{tabular}{|l|r|r|r|r|r|r|} \hline
 &$S_1F_1^SP_{00}^\pi$&$S_1F_2^SP_{00}^K$&$S_1F_3^SP_{00}^\eta$
&$S_2F_1^AP_{00}^\pi$&$S_2F_2^AP_{00}^\eta$\\ \hline
 &&&&& \\
$p$ & $4P_{00}^{\pi}$ &0&${4\over 3}P_{00}^\eta$
& $4P_{00}^\pi$& $-{4\over3}P_{00}^{\eta}$\\
 &&&&& \\
$n$ & $-{4\over 3}P_{00}^{\pi}$ &0&$-{4\over 9}P_{00}^\eta$
&$-4P_{00}^\pi$&${4\over 3}P_{00}^\eta$\\
 &&&&& \\
$\Lambda$ &0& $-{4\over 3}P_{00}^K$
&${4\over 9}P_{00}^\eta$ &0& $-{4\over
3}P_{00}^\eta$\\
 &&&&& \\
$\Sigma^+$ & ${32\over 9}P_{00}^\pi$
&${8\over 9}P_{00}^K$&${8\over 9}
P_{00}^\eta$&0& ${8\over 3}P_{00}^\eta$\\
 &&&&& \\
$\Sigma^0$ & ${8\over 9}P_{00}^\pi$&$-{4\over 9}P_{00}^K$
&${4\over 9}P_{00}^\eta$&0& ${4\over
3}P_{00}^\eta$\\
 &&&&& \\
$\Sigma^0\rightarrow \Lambda$ &0& $-{4\over \sqrt{3}}P_{00}^K$ 
&${4\over 3\sqrt{3}}P_{00}^\eta$
& $-{4\over \sqrt{3}}P_{00}^\pi$&0\\
 &&&&& \\
$\Sigma^-$ & $-{16\over 9}P_{00}^\pi$ 
&$-{16\over 9}P_{00}^K$&0&0&0\\
 &&&&& \\
$\Xi^0$ &0& ${8\over 9}P_{00}^K$ 
&$-{8\over 3}P_{00}^\eta$&0& $-{8\over
3}P_{00}^\eta$\\
 &&&&& \\
$\Xi^-$ &0& $-{16\over 9}P_{00}^K$
&$-{16\over 9}P_{00}^\eta$ & 0&0\\
 &&&&&\\ \hline
 &&&&& \\
$\Delta^{++}$ & $16P_{00}^\pi$&0
&${16\over 3}P_{00}^\eta$ &0&0\\
 &&&&& \\
$\Delta^+\rightarrow p$ & ${8\over 3\sqrt{2}}P_{00}^\pi$
&0&${8\over 9\sqrt{2}}P_{00}^\eta$ &${8\over\sqrt{2}}P_{00}^\pi$&
$-{8\over 3\sqrt{2}}P_{00}^\eta$\\
 &&&&&\\
$\Omega^-$ &0&0& $-{32\over 3}P_{00}^\eta$ & 0&0\\ 
 &&&&& \\ \hline
\end{tabular}
\end{center}
\end{document}